\documentclass[a4paper,11pt]{article}
\usepackage{pos}
\usepackage{cleveref}

\title{Towards symmetric discretization schemes via weak boundary conditions}
\ShortTitle{Towards symmetric discretization schemes}

\author*[a]{Alexander Rothkopf}

\affiliation[a]{Faculty of Science and Technology, University of Stavanger, 4021 Stavanger, Norway}

\emailAdd{alexander.rothkopf@uis.no}

\abstract{The Szymanzik improvement program for gauge theories is most commonly implemented using forward finite difference corrections to the Wilson action. Central symmetric schemes naively applied, suffer from a doubling of degrees of freedom, identical to the well known fermion doubling phenomenon. And while adding a complex Wilson term remedies the problem for fermions, it does not easily transfer to real-valued gauge fields. In this talk I report on recent progress in formulating symmetric discretization schemes for classical actions of simple one-dimensional problems. They avoid doubling by exploiting the weak imposition of initial/boundary conditions. Inspired by recent work in the field of numerical analysis of partial differential equations, I construct a regularized summation-by-parts finite difference operator using boundary data based on affine coordinates. Application to a classical initial value problems with second order derivatives are presented.}

\FullConference{%
The 39th International Symposium on Lattice Field Theory,\\
8th-13th August, 2022,\\
Rheinische Friedrich-Wilhelms-Universität Bonn, Bonn, Germany
}

\begin{document}
\maketitle

\section{Motivation}

Many physical systems of experimental interest are of finite extent, be it the droplet of quark-gluon plasma created in the interior of a relativistic heavy-ion collision \cite{Shen:2020mgh} or a cavity which strongly couples the light field to an electron \cite{frisk2019ultrastrong,forn2019ultrastrong}. In each case weak-coupling methods fail and a lattice field theory evaluation of observables is called for. A finite volume entails loss of translational invariance, similar to when localized sources are explicitly placed in a system, a scenario relevant for the study of e.g. quarkonium bound states in extreme conditions \cite{Rothkopf:2019ipj}. The question I would thus like to address is how to develop improved discretization schemes for systems without translation invariance. 

\begin{figure}[b]
\centering
\includegraphics[scale=0.5]{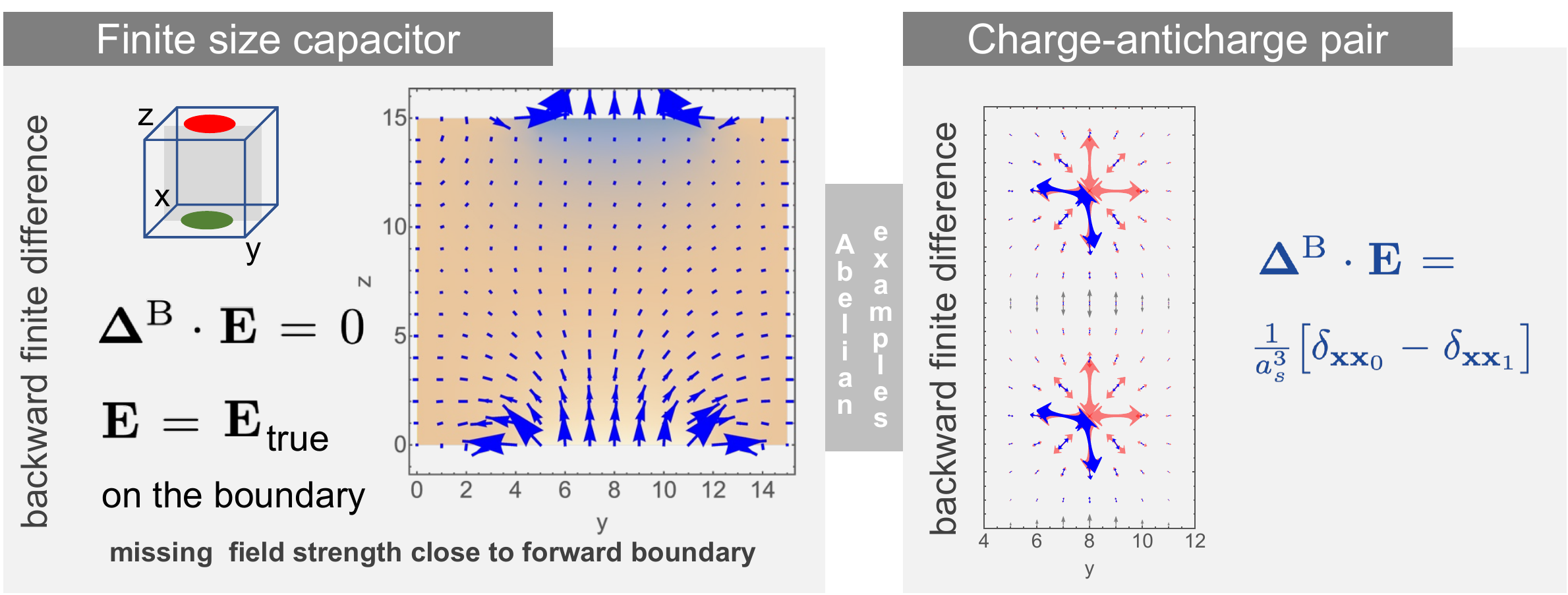}
\caption{(left) field lines from solving Gauss' law in the interior of a capacitor for prescribed values of the true electric field on the boundary (plates). Due to the backward finite difference prescription, field strength is only sustained close to the backward boundary. (right) The field lines arising from solving the BFD Gauss law in the presence of a charge anticharge pair described by point sources (true solution in light red).}\label{fig:motivation}
\end{figure}

The historic starting point for the discretization of lattice gauge theory is Wilson's plaquette action \cite{Wilson:1974sk}, \vspace{-0.5cm}
\begin{align}
&P^{1\times1}_{\mu\nu,x}=U_{\mu,x} U_{\nu,x+a_\mu{\hat\mu}}U^\dagger_{\mu,x+a_\nu{\hat\nu}}U^\dagger_{\nu,x} = e^{i a_\mu a_\nu \tilde F_{\mu\nu,x}}+{\cal O}(a^2),\\
&\tilde F_{\mu\nu}=\Delta^{\rm F}_\mu A_{\nu,x} - \Delta^{\rm F}_\nu A_{\mu,x} + i [ A_{\mu,x},A_{\nu,x}],
\end{align}
which corresponds to a forward finite difference approximation $\Delta^{\rm F}_\mu \phi(x)= (\phi(x+a_\mu \hat\mu) - \phi(x) )/a_\mu$ of the field strength tensor. The associated Gauss law turns out to be described by a backward finite difference (BFD) operator. The limitations of such a BFD discretization are already visible on the level of classical electrodynamics. Inspired by \cite{Yanagihara:2020tvs} I discussed in \cite{Rothkopf:2021jye} two relevant Abelian model systems: a capacitor with a finite charge density on its plates, as well the as a charge-anticharge pair. Prescribing the true values of the electric field on the boundary plates, the BFD discretization $\Delta^B\cdot {\bf E}=0$ manages to sustain field strength in the interior of the capacitor only close to the backward boundary. As shown on the left of \cref{fig:motivation} the field values on the forward facing plate are invisible to the BFD operator. When solving Gauss' law in the presence of a charge anticharge pair (right panel of \cref{fig:motivation}) and computing the field lines in a gauge invariant fashion by diagonalizing the stress tensor (blue arrows), one sees that they show a significant imbalance towards the backward direction. 

The gauge actions deployed in the lattice QCD community today are improved, in the sense of the Szymanzik program \cite{symanzik1983continuum}. To accelerate the approach to the continuum limit, i.e. to reduce discretization artifacts in the simulated correlators, the plaquettes of the Wilson action are amended by loops with larger than unit area. Most common correction terms are the forward rectangles. And while these terms indeed reduce the lattice spacing dependence of the simulation they do not realize a symmetric discretization of the field strength around charges. Recovering the Gauss law in the continuum limit therefore becomes non-trivial.

To realize a genuine symmetric discretization scheme, I proposed in \cite{Rothkopf:2021jye} to use instead the $2\times 2$ plaquette centered around the nodes of the grid \vspace{-0.2cm}
\begin{align}
\nonumber P^{2\times2}_{\mu\nu,x}=& \bar U_{\mu,x-a\hat\mu-a\hat\nu}\bar U_{\mu,x-a\hat\nu}\bar U_{\nu,x+a\hat\mu-a\hat\nu}\bar U_{\nu,x+a\hat\mu} \bar U^\dagger_{\mu,x+a\hat\nu}\bar U^\dagger_{\mu,x-a\hat\mu+a\hat\nu}\bar U^\dagger_{\nu,x-a\hat\mu}\bar U^\dagger_{\nu,x-a\hat\mu-a\hat\nu}\\
\nonumber =&{\rm exp}\big[ 4ig a_\mu a_\nu \bar F_{\mu\nu,x}\big] + {\cal O}(a^3), \quad \bar F_{\mu\nu,x}= {\bf \Delta}^{\rm C}_\mu A_{\nu,x} - {\bf \Delta}^{\rm C}_\nu A_{\mu,x}+i[A_{\mu,x},A_{\nu,x}]
\end{align}
which realizes a central finite difference discretization $\Delta^{\rm C}_\mu \phi(x)= (\phi(x+a_\mu \hat\mu) - \phi(x-a_\mu \hat\mu) )/2a_\mu$ of the field strength. This plaquette differs from the usual clover leaf prescription in that it corresponds to the product of four $1\times 1$ plaquettes and thus remains within the gauge group (see also its relation to stabilized Wilson fermions \cite{Francis:2022hyr}).

\begin{figure}
\centering
\includegraphics[scale=0.75]{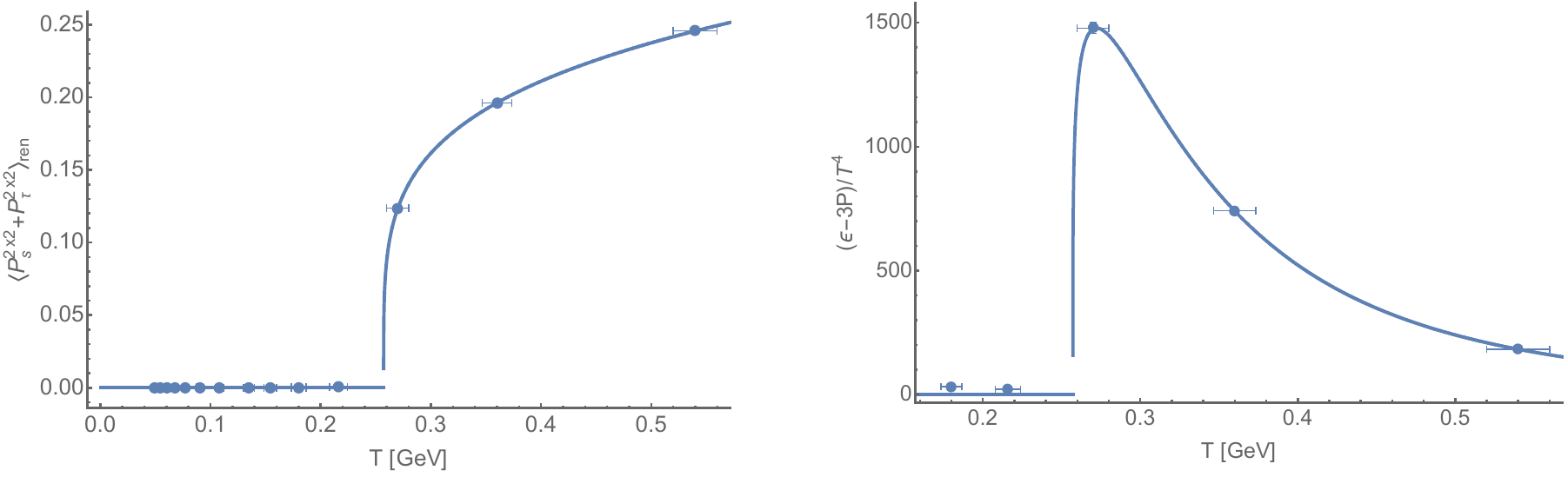}
\caption{(left) Renormalized sum of the spatial and temporal $P^{2\times2}$ plaquettes that enter the computation of the trace anomaly in simulations with $S^{2\times 2}$. (right) Values of the trace anomaly from naively implemented $S^{2\times 2}$, which is larger than the correct value by a factor of $4\times8\times15$.}\label{fig:doublerlat}\end{figure}

First simulations \cite{Horowitz:2021dmr} based on the action $S^{2\times 2}=\sum_{x} a_ta_s^3\Big[ \frac{2}{16 a_t^2a_s^2}\sum_i{\rm ReTr}\big[ 1-P^{2\times2}_{0i,x}\big] - \frac{1}{16 a_s^4}\sum_{ij} {\rm ReTr}\big[ 1-P^{2 \times2}_{ij,x}\big] \Big]$ however revealed (see \cref{fig:doublerlat}) that such a symmetric discretization scheme for gauge fields suffers from \textit{bosonic doublers}, similar to the ones encountered when using symmetric finite differences in the discretization of fermion fields. I.e. the values of the trace anomaly in the right panel of \cref{fig:doublerlat} are larger than the correct result by a factor of $4\times8\times15$ (c.f. \cite{umeda2009fixed}), indicating that for each physical mode, fifteen unphysical degrees of freedom propagate in the interior.

\section{Regularization with boundary data}

\begin{figure}
\centering
\includegraphics[scale=0.75]{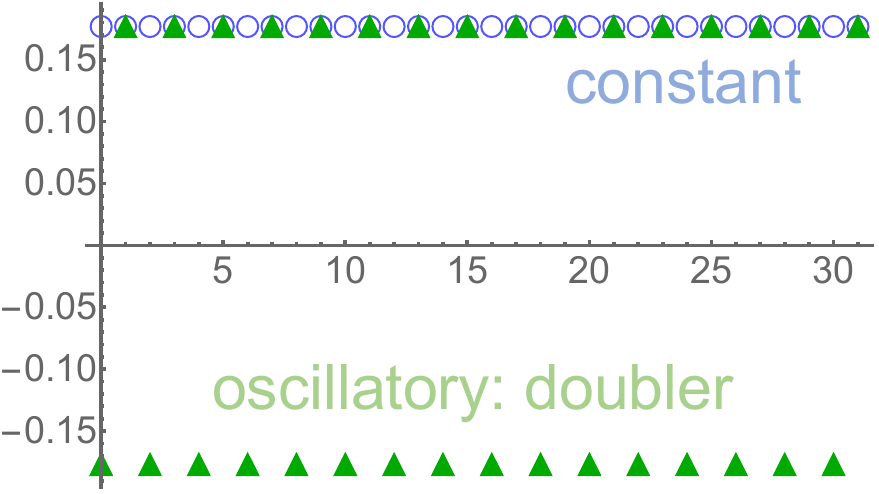}\hspace{0.5cm}
\includegraphics[scale=0.6]{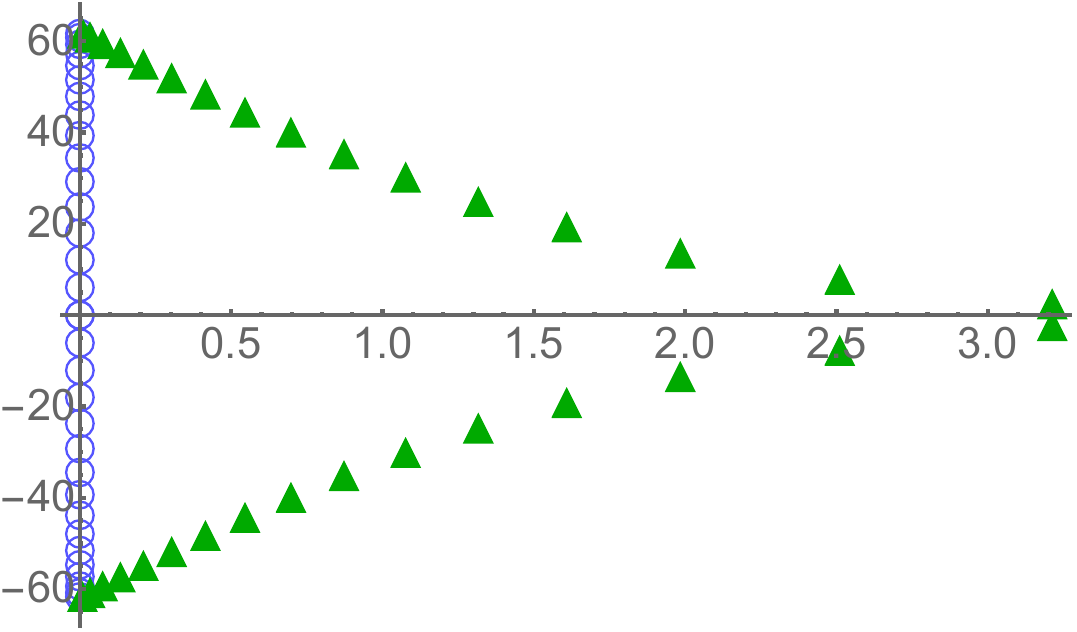}
\caption{(left) The degenerate eigenfunctions associated with the zero eigenvalues of $D^C$ (blue circles) and with its transpose $(D^C)^t$ (green triangles). (right) Eigenvalue spectrum of the operator $D^C$ with exact zero modes (blue circles) and of the regularized operator $\tilde D$ (green triangles).}\label{fig:eigenspecs}
\end{figure}

Let us illustrate the doubling problem in coordinate space in one dimension, using the following finite difference operator $D^C$ with symmetric stencil in the interior
\begin{align} 
D^{\rm C}=\frac{1}{\Delta x}\left[ \begin{array}{cccc} -1 & 1 & 0 & 0 \\ -\frac{1}{2} &0 &\frac{1}{2} & 0 \\ 0 & -\frac{1}{2} &0 &\frac{1}{2} \\ 0&0&-1&1 \end{array}\right],\qquad \tilde D = D^C+\kappa \frac{\Delta x}{2\Delta x^2}\left[ \begin{array}{cccc} \ddots & 1 & 0 & 0 \\ 1 &-2 &1 & 0 \\ 0 & 1 &-2 &1 \\ 0&0&1&\ddots \end{array}\right].\label{eq:DandDreg}
\end{align}
The operator $D^C$ is a so-called summation by parts operator, as it mimics accurately integration by parts in the discrete setting. (see e.g. \cite{fernandez2014review} and for the related discussion of the momentum operator on finite domains see \cite{Al-Hashimi:2021uhu,Albrecht:2022sdd}). If we compute its spectrum, we find that it features two zero eigenvalues. They are associated with degenerate eigenfunctions that turn out to be the constant function (blue circles in the left panel of \cref{fig:eigenspecs}). The doublers are hiding in the eigenfunctions of the transpose $(D^C)^t$ and are indeed the maximally oscillating function on the grid also with eigenvalue zero (green triangles).

In the numerical analysis literature several regularization approaches are discussed. One of them, called the \textit{upwind modification} ($\kappa=1$ in \cref{eq:DandDreg} ), adds a symmetric second derivative to $D^C$ with one extra power of the grid spacing $\Delta x$. This term vanishes in the continuum limit and does not affect the defining property of the finite difference, i.e. $\tilde D {\bf x}^r = r {\bf x}^{r-1}$ for $r$ smaller than the order of $D^C$.  At the same time this modification destroys the symmetry of the interior stencil and reverts the derivative to the naive forward form. It was Wilson's seminal contribution \cite{Wilson:1974sk} to realize that for complex valued fermionic fields the higher order derivative can be added with an imaginary prefactor ($\kappa=i$ in \cref{eq:DandDreg}) which retains the symmetry of both terms while lifting the zero modes. This is the celebrated \textit{Wilson term}. In case of bosonic gauge fields, which must remain real-valued, a similar modification however is not possible.

Taking inspiration from the computational fluid dynamics community, I propose to use another lever to regularize the bosonic doubler problem: boundary values. In a finite system boundary data are physical information and otherwise can be chosen at convenience. In the past, boundary conditions were mostly implemented in the strong sense, i.e one replaces the degrees of freedom on the boundary apriori with the prescribed values. Alternatively one may consider the \textit{weak treatment} of boundary conditions, which acknowledges that the boundary conditions only need to be realized as accurately as the rest of the discretization. In turn one can introduce boundary information through penalty terms, which offer new opportunities for regularization. This approach, known as simultaneous approximation terms (SAT) \cite{carpenter1994time} is well established in the treatment of classical ODEs and PDEs \cite{lundquist2014sbp}.

Consider the continuum problem of solving the ODE $u'(x)=g(x)$ with boundary condition $u(0)=u_0$. In discretized form it can be formulated as
\begin{align}
D{\bf u}={\bf g} + \frac{1}{\Delta x}E_0\big({\bf u}-{\bf u}_0\big) \quad\rightarrow\quad \tilde D=D-\frac{1}{\Delta x}E_0 \quad\rightarrow\quad \quad \tilde D{\bf u}={\bf g}-\frac{1}{\Delta x}E_0{\bf u}_0.\label{eq:ODEex}
\end{align}
where we use the notation ${\bf u}_k=u(k \Delta x)$ and introduce a penalty term that refers to the boundary values via the projection matrix $E_0={\rm diag}[1,0,\ldots]$ and ${\bf u}_0=\{u_0,0,\ldots\}$. As the lattice spacing is reduced the penalty increases and the boundary conditions will be more strictly fulfilled. The form of the penalty term invites us to absorb its homogeneous part into a redefined difference operator $\tilde D$. In the right panel of \cref{fig:eigenspecs} we show the effect of this redefinition on the eigenvalues of the finite difference operator. The blue points represent the eigenvalues of $D$, which feature two exact zeros connected to the doubler mode. The green points on the other hand denote the eigenvalues of $\tilde D$ where all zero modes have been lifted. Thus $\tilde D$ is an invertible operator leading to a unique solution of the system of equations for ${\bf u}$  in the right-most term in \cref{eq:ODEex}.

On the lattice we need to incorporate this regularization in the action of the system. The difficulty here lies in the fact that now we do not have an equal sign to move the penalty term around, as we did in \cref{eq:ODEex}. Together with J. Nordst\"om, I recently proposed a solution by incorporating the penalty term as a whole in the definition of the finite difference operator using \textit{affine coordinates}. Take as example the continuum action $S=\int dx u'(x)u'(x)$ with some boundary condition $u(0)=u_0$. Implementing the quadrature of the integral using a matrix (e.g. $H=\Delta x {\rm diag}[\frac{1}{2},1,\ldots,1,\frac{1}{2}]$ for the trapezoid rule) we can write $S\approx (D{\bf u})^t H D{\bf u}$. Our proposal is to define a new regularized finite difference operator in which the boundary penalty term is included
\begin{align}
\bar D {\bf u}=D{\bf u}+H^{-1}E_0\big({\bf u}-{\bf u}_0\big), \qquad \bar D =\frac{1}{\Delta x}\left[ \begin{array}{ccccc} -1 + 2 & 1 & 0 & 0 &-2 u_0 \\ -\frac{1}{2} &0 &\frac{1}{2} & 0 & 0\\ 0 & -\frac{1}{2} &0 &\frac{1}{2} &0 \\ 0&0&-1&1 &0 \\ 0&0&0&0&1\end{array}\right].\label{eq:DregAff}
\end{align}
The last term involving ${\bf u}_0$ is nothing but a shift, which can be conveniently included in matrix form when amending the matrix of $\bar D$ by one row and one column, placing the value one in the lower right corner and filling the extra column on the right with the values of the shift (see term on the right in \cref{eq:DregAff}). For consistency all vectors corresponding to discretized functions are also amended with one more entry of the value one $\tilde {\bf u}=\{{\bf u},1\}$. The absorption of the boundary term into $\bar D$ has a similar effect on its eigenvalues as we observed for the regularized $\tilde D$ previously. Both zero modes are lifted. For $\bar D$ expressed in affine coordinates we find that there exists a single purely real eigenvalue of value one, while all others come in complex conjugate pairs. This eigenvalue is now associated with the zero function ${\bf u}^{(0)}_{\rm aff}=\{0,\ldots,0,1\}$ in the spectrum of $\bar D^t$, from which the maximally oscillating doubler mode has been deleted.

The inclusion of boundary data as shown here, constitutes a novel regularization procedure for symmetric discretization schemes. It is applicable also to purely real field degrees of freedom, in contrast to the Wilson term, which requires complex valued functions. 

\section{Application to simple initial value problems}

As a first application of the regularization procedure it has been used in \cite{Rothkopf:2022zfb} to develop a novel discretization prescription for classical initial value problems (IVP), based on a variational principle. The long-term goal of this line of study is to realize genuine real-time simulations of quantum fields on the lattice (see e.g. our work on complex Langevin \cite{Alvestad:2021hsi} ). However as an intermediate time goal I see the realization of gauge invariant simulations of the real-time dynamics of classical lattice gauge theory. Today these simulations are based on Hamilton's equations of motion, which require a choice of gauge. The first modest step in this direction I am going to discuss here is the development of a variational solver for initial value problems in classical point mechanics.
\begin{figure}
\centering
\includegraphics[scale=0.5]{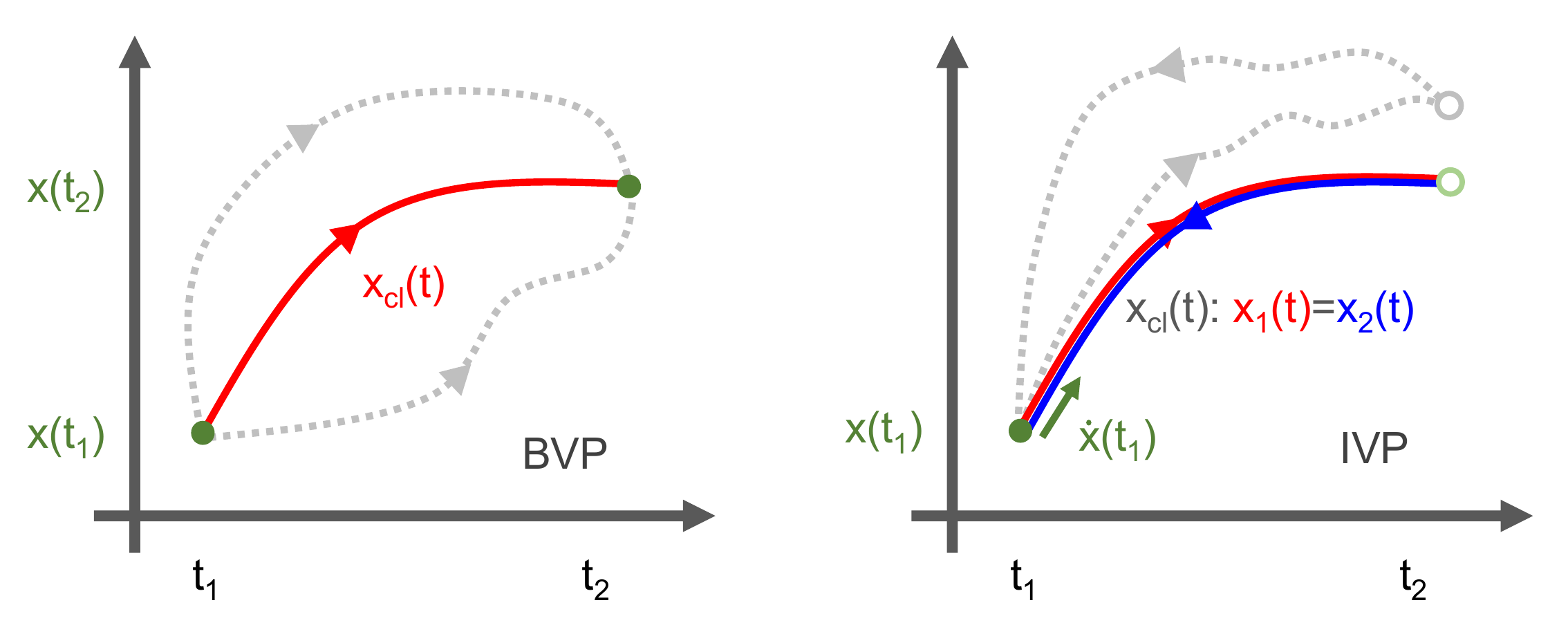}
\caption{Sketches of (left) the conventional boundary value formulation of the variational principle in classical mechanics and (right) the variational principle adapted to initial value problems, which requires a doubling of degrees of freedom.}\label{fig:BVPIVPvar}
\end{figure}
The challenge one faces with the construction of variational solvers for IVPs is the fact that conventionally the variational problem is formulated as a boundary value problem (see left panel of \cref{fig:BVPIVPvar}). This is unsatisfactory, as the position of the system at final time $t_2$ needs to be provided apriori, an information that is not available in a genuine IVP. As was shown in \cite{galley_classical_2013} a variational principle can be established when introducing a doubling of the degrees of freedom, leading in essence to a double shooting method. In essence \cite{galley_classical_2013} discusses the classical limit of the Schwinger-Keldysh real-time contour prescription, where in addition to degrees of freedom on the forward branch $x_1$ also $x_2$ on a backward time branch are considers. The latter are assigned a negative weight in the system action.
\begin{align}
    {\cal S}_{\rm IVP}[x_1(t),\dot x_1(t),x_2(t),\dot x_2(t)]&=\int_{t_1}^{t_2} dt\Big( {\cal L}[x_1(t),\dot x_1(t)] - {\cal L}[x_2(t),\dot x_2(t)] \Big),\\&=\int_{t_1}^{t_2} dt \Big( {\rm L}[x_1(t),\dot x_1(t),x_2(t),\dot x_2(t)] \Big).\label{eq:SIVP}
\end{align}
Let us introduce the coordinates $x_+=(x_1+x_2)/2$ and $x_-=x_1-x_2$, so that the variation of the system action can be expressed in the form
\begin{align}
    \delta {\cal S} = \int dt \Big( \Big\{ \frac{\partial {\rm L}}{\partial x_1}- \frac{d}{dt}\frac{\partial {\rm L}}{\partial \dot x_1}   \Big\}\delta x_1 -\Big\{ \frac{\partial {\rm L}}{\partial x_2}- \frac{d}{dt}\frac{\partial {\rm L}}{\partial \dot x_2}   \Big\}\delta x_2\Big) + \left. \Big[ \frac{\partial {\rm L}}{\partial \dot x_1} \delta x_1 \Big]\right|_{t_1}^{t_2} - \left.\Big[ \frac{\partial {\rm L}}{\partial \dot x_2} \delta x_2 \Big]\right|_{t_1}^{t_2}  \label{eq:equivEL}.
\end{align}
Now if one enforces that the value of the paths and their derivatives agree at the final time step $t_2$ one can show that the critical point of the action is equivalent to the solution of the Euler-Lagrange equations. I.e. even though the value of the classical path at the final time step $t_2$ is not fixed to a certain value, the contributions from $x_1(t_2)$ and $x_2(t_2)$ are designed such that they cancel correctly the boundary terms arising there. The stationarity condition arising from this variational principle is most concisely formulated as $\left. \frac{\delta S_{\rm IVP}[x_{\pm}]}{\delta x_-}\right|_{x_-=0,x_+=x_{\rm class}}=0$. The above derivation did not make any reference to quantum field theory, but reassuringly reproduces the result of \cite{berges2007quantum} where the classical limit of the Schwinger-Keldysh formalism was investigated.

\begin{figure}
\centering
\includegraphics[scale=0.2]{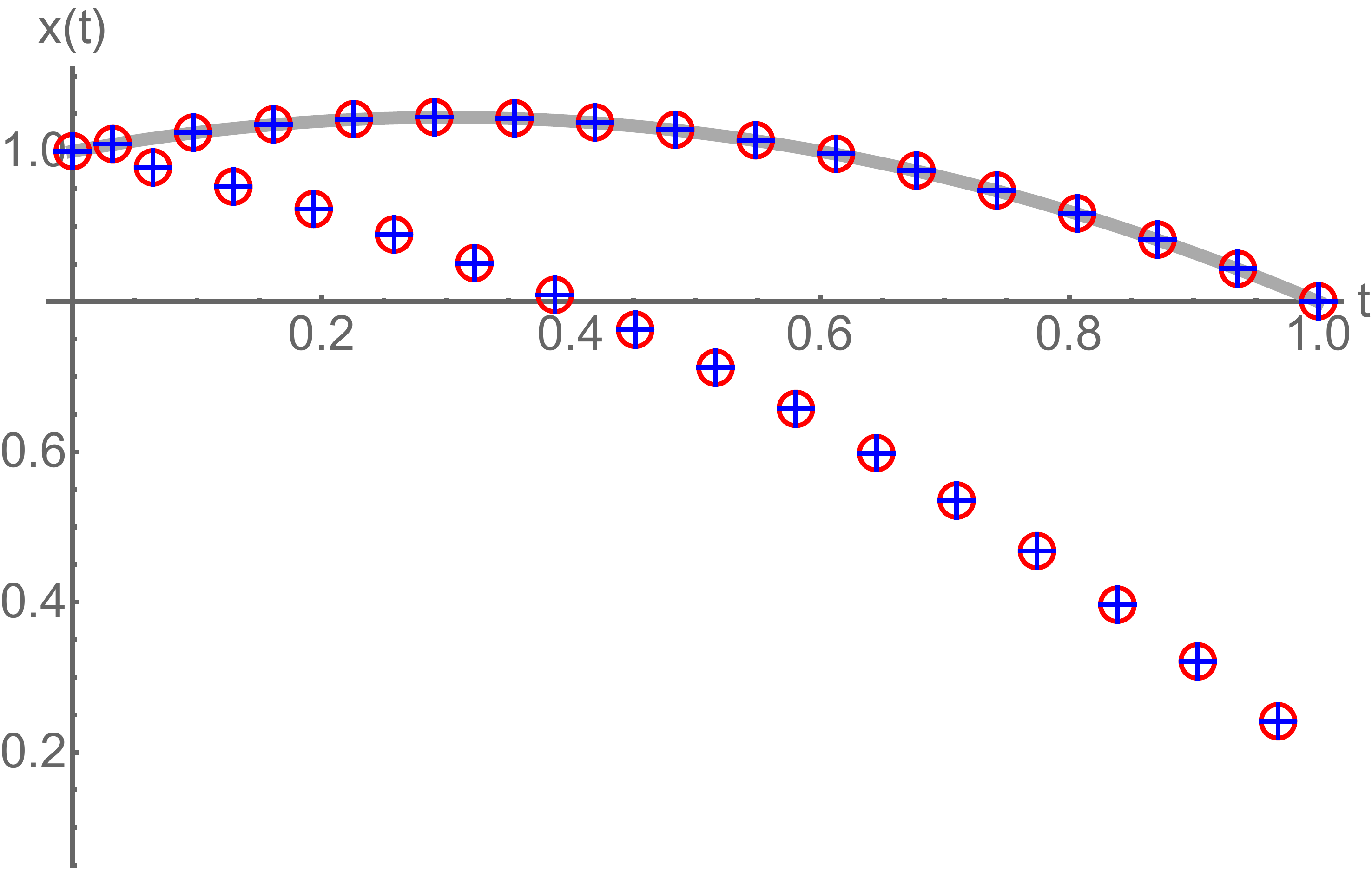}\hspace{0.7cm}
\includegraphics[scale=0.2]{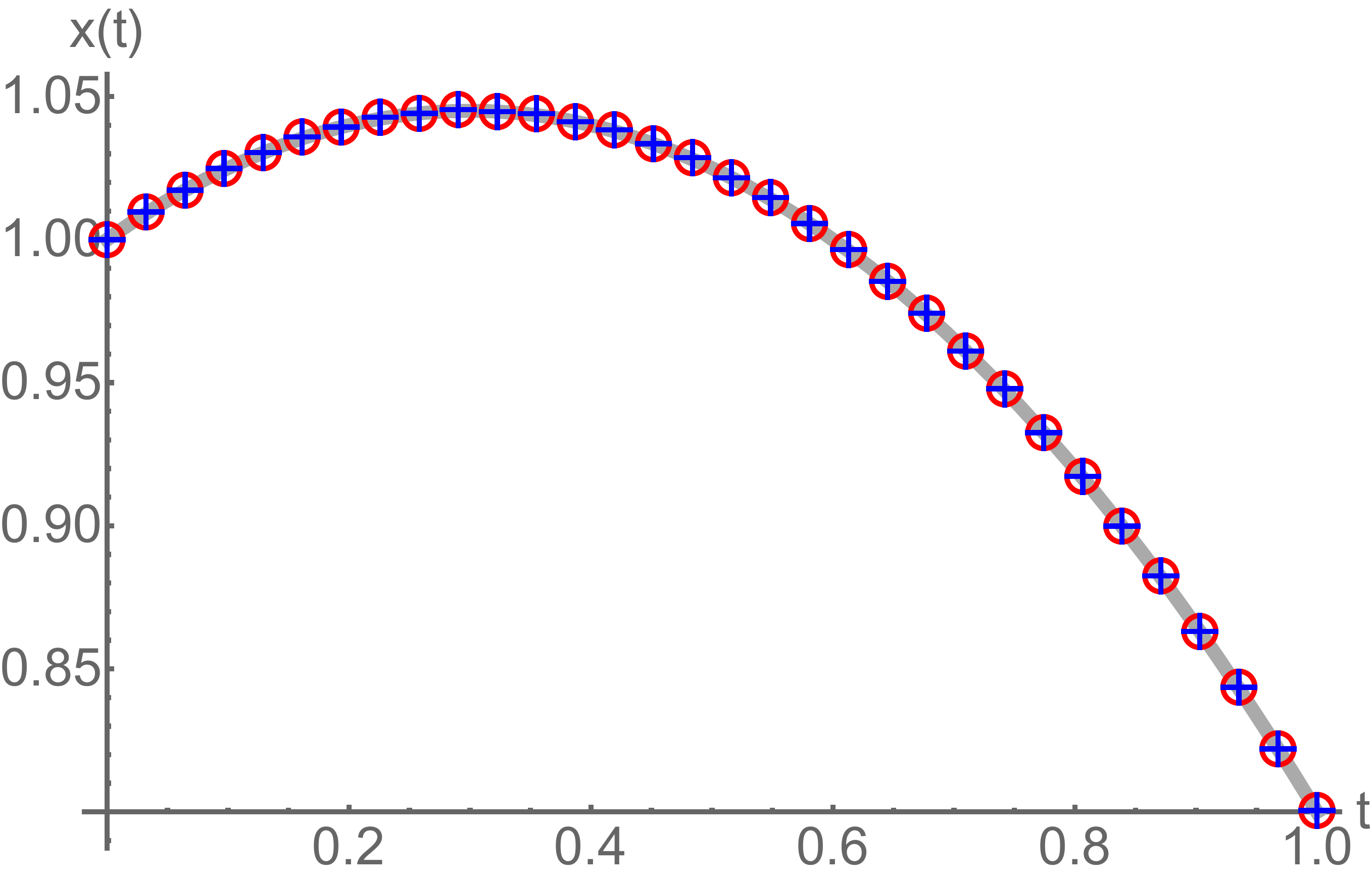}
\caption{(left) Solution of the IVP of \cref{eq:IVPfunc} with a naive SBP operator which suffers from the occurrence of doublers. (right) Solution of the IVP of \cref{eq:IVPfuncreg} with the regularized SBP operator in affine coordinates which avoids doublers. Note that the solutions for ${\bf x}_1$ (crosses) and ${\bf x}_2$ (circles) agree as required by the stationarity condition.}\label{fig:numIVP}
\end{figure}

We set out to discretize and solve the variational principle for a simple IVP, the point particle in a constant gravitational field with continuum action ${\cal S}=\int\,dt\Big(\frac{1}{2}m\dot x^2(t)-mgx(t)\Big)$ and arbitrarily chosen initial conditions $x(0)=1$ and $\dot x(0)=0.3$. According to \cref{eq:SIVP} we must double the degrees of freedom, hence we introduce ${\bf x}_1$ and ${\bf x}_2$. The naive discretization of the action with quadrature matrix $H$ and the naive SBP finite difference operator $D^C$ introduced in \cref{eq:DandDreg} reads 
\begin{align}
&{\cal S}_{\rm IVP}= \Big\{  \frac{1}{2} (D^C{\bf x}_1)^{\rm T} {H} (D^C{\bf x}_1) - g {1}^{\rm T} {H} {\bf x}_1\Big\} - \Big\{\frac{1}{2} (D^C{\bf x}_2)^{\rm T} {H} (D^C{\bf x}_2) - g {1}^{\rm T}  {H} {\bf x}_2 \Big\}\label{eq:IVPfunc}\\
\nonumber &+ \lambda_1 (x_1(0)-x_i) + \lambda_2((D^C{\bf x}_1)(0)-\dot x_i) + \lambda_3 (x_1(N_t)-x_2(N_t)) + \lambda_4 ( (D^C{\bf x}_1)(N_t)-(D^C{\bf x}_2)(N_t) ).
\end{align}
In order to locate the critical point of the action under the constraints that initial conditions shall be fulfilled and that the paths ${\bf x}_1$ and ${\bf x}_2$ shall agree at the last time point we have added four Langrange multipliers $\lambda_i$. Using $N_t=32$ steps and a time extent of $t_2=1$, we carry out a numerical minimization of the expression in \cref{eq:IVPfunc} which leads to the result shown in the left panel of \cref{fig:numIVP}. We find that the obtained solution (crosses for ${\bf x}_1$, circles for ${\bf x}_2$)  fulfills the requirement ${\bf x}_1={\bf x}_2$ of the stationarity condition. One can however clearly see that only a subset of the solution lies on the correct solution (gray solid line) and that a significant portion of the points on the paths form an artificial oscillatory pattern. From our investigation of the spectrum of $D^C$ this is not surprising, as the maximally oscillating zero mode, the doubler, has simply contaminated the result.

If instead we use the finite difference operator $\bar D$ in affine coordinates, which is regularized using the initial value data, together with the corresponding paths $\bar {\bf x}_{1,2}$ and quadrature matrix $\bar H$ in affine coordinates, we end up with the following  expression
\begin{align}
&{\cal S}^{\rm reg}_{\rm IVP}= \Big\{  \frac{1}{2} (\bar D\bar{\bf x}_1)^{\rm T} {\bar H} (\bar D\bar{\bf x}_1) - g {1}^{\rm T} {H} {\bf x}_1\Big\} - \Big\{\frac{1}{2} (\bar D\bar {\bf x}_2)^{\rm T} {\bar H} (\bar D\bar {\bf x}_2) - g {1}^{\rm T}  {H} {\bf x}_2 \Big\}\label{eq:IVPfuncreg}\\
\nonumber &+ \lambda_1 (x_1(0)-x_i) + \lambda_2((D^C{\bf x}_1)(0)-\dot x_i) + \lambda_3 (x_1(N_t)-x_2(N_t)) + \lambda_4 ( (D^C{\bf x}_1)(N_t)-(D^C{\bf x}_2)(N_t) ).
\end{align}
whose critical trajectory is shown on the right of \cref{fig:numIVP}. The regularization has successfully avoided the occurrence of doublers and manages to bring the solution close to the correct classical trajectory. Of course the accuracy of the solution is limited by the accuracy of the discretization used for the finite difference operator $\bar D$.  Systematic prescriptions for the construction of higher order SBP operators exist and the regularization in affine coordinates is independent of the form of $D$, i.e. it can be applied straight forwardly to  higher order operators too. We have checked that the classical trajectory found via this variational approach correctly approaches the continuum limit under grid refinement (for a detailed analysis see \cite{Rothkopf:2022zfb}).

\section{Conclusion}
The physics of strongly correlated fields in finite systems and in the presence of explicit sources requires discretization schemes in the absence of translation invariance. A proposal to deploy the centrally symmetric $2\times2$ plaquette revealed the occurrence of \textit{bosonic doublers} in such symmetric discretization schemes. The challenge lies in the fact that the Wilson term regularization, successful for complex valued fermion fields is not applicable for real-valued gauge fields. Instead I propose to use the weak imposition of boundary data as an alternative regularization mechanism, which can be straight forwardly implemented when expressing finite difference operators in affine coordinates. As a first step, the efficacy of this type of regularization has been demonstrated in a variational solver for classical initial value problems. While unregularized finite difference operators lead to solutions for the classical trajectory that are contaminated by doubler modes, the regularized operator successfully avoids the occurrence of the bosonic doublers. Extension of the variational approach to higher dimensions is work in progress.

\section{Acknowledgments}
A.R. is supported by the Research Council of Norway under the FRIPRO Young Research Talent grant 286883.

\bibliographystyle{apsrev4-1} 
\bibliography{references}

\begin{thebibliography}{20}%
\makeatletter
\providecommand \@ifxundefined [1]{%
 \@ifx{#1\undefined}
}%
\providecommand \@ifnum [1]{%
 \ifnum #1\expandafter \@firstoftwo
 \else \expandafter \@secondoftwo
 \fi
}%
\providecommand \@ifx [1]{%
 \ifx #1\expandafter \@firstoftwo
 \else \expandafter \@secondoftwo
 \fi
}%
\providecommand \natexlab [1]{#1}%
\providecommand \enquote  [1]{``#1''}%
\providecommand \bibnamefont  [1]{#1}%
\providecommand \bibfnamefont [1]{#1}%
\providecommand \citenamefont [1]{#1}%
\providecommand \href@noop [0]{\@secondoftwo}%
\providecommand \href [0]{\begingroup \@sanitize@url \@href}%
\providecommand \@href[1]{\@@startlink{#1}\@@href}%
\providecommand \@@href[1]{\endgroup#1\@@endlink}%
\providecommand \@sanitize@url [0]{\catcode `\\12\catcode `\$12\catcode
  `\&12\catcode `\#12\catcode `\^12\catcode `\_12\catcode `\%12\relax}%
\providecommand \@@startlink[1]{}%
\providecommand \@@endlink[0]{}%
\providecommand \url  [0]{\begingroup\@sanitize@url \@url }%
\providecommand \@url [1]{\endgroup\@href {#1}{\urlprefix }}%
\providecommand \urlprefix  [0]{URL }%
\providecommand \Eprint [0]{\href }%
\providecommand \doibase [0]{http://dx.doi.org/}%
\providecommand \selectlanguage [0]{\@gobble}%
\providecommand \bibinfo  [0]{\@secondoftwo}%
\providecommand \bibfield  [0]{\@secondoftwo}%
\providecommand \translation [1]{[#1]}%
\providecommand \BibitemOpen [0]{}%
\providecommand \bibitemStop [0]{}%
\providecommand \bibitemNoStop [0]{.\EOS\space}%
\providecommand \EOS [0]{\spacefactor3000\relax}%
\providecommand \BibitemShut  [1]{\csname bibitem#1\endcsname}%
\let\auto@bib@innerbib\@empty
\bibitem [{\citenamefont {Shen}\ and\ \citenamefont
  {Yan}(2020)}]{Shen:2020mgh}%
  \BibitemOpen
  \bibfield  {author} {\bibinfo {author} {\bibfnamefont {C.}~\bibnamefont
  {Shen}}\ and\ \bibinfo {author} {\bibfnamefont {L.}~\bibnamefont {Yan}},\
  }\href {\doibase 10.1007/s41365-020-00829-z} {\bibfield  {journal} {\bibinfo
  {journal} {Nucl. Sci. Tech.}\ }\textbf {\bibinfo {volume} {31}},\ \bibinfo
  {pages} {122} (\bibinfo {year} {2020})},\ \Eprint
  {http://arxiv.org/abs/2010.12377} {arXiv:2010.12377 [nucl-th]} \BibitemShut
  {NoStop}%
\bibitem [{\citenamefont {Frisk~Kockum}\ \emph {et~al.}(2019)\citenamefont
  {Frisk~Kockum}, \citenamefont {Miranowicz}, \citenamefont {De~Liberato},
  \citenamefont {Savasta},\ and\ \citenamefont {Nori}}]{frisk2019ultrastrong}%
  \BibitemOpen
  \bibfield  {author} {\bibinfo {author} {\bibfnamefont {A.}~\bibnamefont
  {Frisk~Kockum}}, \bibinfo {author} {\bibfnamefont {A.}~\bibnamefont
  {Miranowicz}}, \bibinfo {author} {\bibfnamefont {S.}~\bibnamefont
  {De~Liberato}}, \bibinfo {author} {\bibfnamefont {S.}~\bibnamefont
  {Savasta}}, \ and\ \bibinfo {author} {\bibfnamefont {F.}~\bibnamefont
  {Nori}},\ }\href@noop {} {\bibfield  {journal} {\bibinfo  {journal} {Nature
  Reviews Physics}\ }\textbf {\bibinfo {volume} {1}},\ \bibinfo {pages} {19}
  (\bibinfo {year} {2019})}\BibitemShut {NoStop}%
\bibitem [{\citenamefont {Forn-D{\'\i}az}\ \emph {et~al.}(2019)\citenamefont
  {Forn-D{\'\i}az}, \citenamefont {Lamata}, \citenamefont {Rico}, \citenamefont
  {Kono},\ and\ \citenamefont {Solano}}]{forn2019ultrastrong}%
  \BibitemOpen
  \bibfield  {author} {\bibinfo {author} {\bibfnamefont {P.}~\bibnamefont
  {Forn-D{\'\i}az}}, \bibinfo {author} {\bibfnamefont {L.}~\bibnamefont
  {Lamata}}, \bibinfo {author} {\bibfnamefont {E.}~\bibnamefont {Rico}},
  \bibinfo {author} {\bibfnamefont {J.}~\bibnamefont {Kono}}, \ and\ \bibinfo
  {author} {\bibfnamefont {E.}~\bibnamefont {Solano}},\ }\href@noop {}
  {\bibfield  {journal} {\bibinfo  {journal} {Reviews of Modern Physics}\
  }\textbf {\bibinfo {volume} {91}},\ \bibinfo {pages} {025005} (\bibinfo
  {year} {2019})}\BibitemShut {NoStop}%
\bibitem [{\citenamefont {Rothkopf}(2020)}]{Rothkopf:2019ipj}%
  \BibitemOpen
  \bibfield  {author} {\bibinfo {author} {\bibfnamefont {A.}~\bibnamefont
  {Rothkopf}},\ }\href {\doibase 10.1016/j.physrep.2020.02.006} {\bibfield
  {journal} {\bibinfo  {journal} {Phys. Rept.}\ }\textbf {\bibinfo {volume}
  {858}},\ \bibinfo {pages} {1} (\bibinfo {year} {2020})},\ \Eprint
  {http://arxiv.org/abs/1912.02253} {arXiv:1912.02253 [hep-ph]} \BibitemShut
  {NoStop}%
\bibitem [{\citenamefont {Wilson}(1974)}]{Wilson:1974sk}%
  \BibitemOpen
  \bibfield  {author} {\bibinfo {author} {\bibfnamefont {K.~G.}\ \bibnamefont
  {Wilson}},\ }\href {\doibase 10.1103/PhysRevD.10.2445} {\bibfield  {journal}
  {\bibinfo  {journal} {Phys. Rev. D}\ }\textbf {\bibinfo {volume} {10}},\
  \bibinfo {pages} {2445} (\bibinfo {year} {1974})}\BibitemShut {NoStop}%
\bibitem [{\citenamefont {Yanagihara}\ \emph {et~al.}(2020)\citenamefont
  {Yanagihara}, \citenamefont {Kitazawa}, \citenamefont {Asakawa},\ and\
  \citenamefont {Hatsuda}}]{Yanagihara:2020tvs}%
  \BibitemOpen
  \bibfield  {author} {\bibinfo {author} {\bibfnamefont {R.}~\bibnamefont
  {Yanagihara}}, \bibinfo {author} {\bibfnamefont {M.}~\bibnamefont
  {Kitazawa}}, \bibinfo {author} {\bibfnamefont {M.}~\bibnamefont {Asakawa}}, \
  and\ \bibinfo {author} {\bibfnamefont {T.}~\bibnamefont {Hatsuda}},\ }\href
  {\doibase 10.1103/PhysRevD.102.114522} {\bibfield  {journal} {\bibinfo
  {journal} {Phys. Rev. D}\ }\textbf {\bibinfo {volume} {102}},\ \bibinfo
  {pages} {114522} (\bibinfo {year} {2020})},\ \Eprint
  {http://arxiv.org/abs/2010.13465} {arXiv:2010.13465 [hep-lat]} \BibitemShut
  {NoStop}%
\bibitem [{\citenamefont {Rothkopf}(2021)}]{Rothkopf:2021jye}%
  \BibitemOpen
  \bibfield  {author} {\bibinfo {author} {\bibfnamefont {A.}~\bibnamefont
  {Rothkopf}},\ }\href@noop {} {\  (\bibinfo {year} {2021})},\ \Eprint
  {http://arxiv.org/abs/2102.08616} {arXiv:2102.08616 [hep-lat]} \BibitemShut
  {NoStop}%
\bibitem [{\citenamefont {Symanzik}(1983)}]{symanzik1983continuum}%
  \BibitemOpen
  \bibfield  {author} {\bibinfo {author} {\bibfnamefont {K.}~\bibnamefont
  {Symanzik}},\ }\href@noop {} {\bibfield  {journal} {\bibinfo  {journal}
  {Nuclear Physics B}\ }\textbf {\bibinfo {volume} {226}},\ \bibinfo {pages}
  {187} (\bibinfo {year} {1983})}\BibitemShut {NoStop}%
\bibitem [{\citenamefont {Francis}\ \emph {et~al.}(2022)\citenamefont
  {Francis}, \citenamefont {Cuteri}, \citenamefont {Fritzsch}, \citenamefont
  {Pederiva}, \citenamefont {Rago}, \citenamefont {Schindler}, \citenamefont
  {Walker-Loud},\ and\ \citenamefont {Zafeiropoulos}}]{Francis:2022hyr}%
  \BibitemOpen
  \bibfield  {author} {\bibinfo {author} {\bibfnamefont {A.~S.}\ \bibnamefont
  {Francis}}, \bibinfo {author} {\bibfnamefont {F.}~\bibnamefont {Cuteri}},
  \bibinfo {author} {\bibfnamefont {P.}~\bibnamefont {Fritzsch}}, \bibinfo
  {author} {\bibfnamefont {G.}~\bibnamefont {Pederiva}}, \bibinfo {author}
  {\bibfnamefont {A.}~\bibnamefont {Rago}}, \bibinfo {author} {\bibfnamefont
  {A.}~\bibnamefont {Schindler}}, \bibinfo {author} {\bibfnamefont
  {A.}~\bibnamefont {Walker-Loud}}, \ and\ \bibinfo {author} {\bibfnamefont
  {S.}~\bibnamefont {Zafeiropoulos}},\ }\href {\doibase 10.22323/1.396.0118}
  {\bibfield  {journal} {\bibinfo  {journal} {PoS}\ }\textbf {\bibinfo {volume}
  {LATTICE2021}},\ \bibinfo {pages} {118} (\bibinfo {year} {2022})},\ \Eprint
  {http://arxiv.org/abs/2201.03874} {arXiv:2201.03874 [hep-lat]} \BibitemShut
  {NoStop}%
\bibitem [{\citenamefont {Horowitz}\ and\ \citenamefont
  {Rothkopf}(2022)}]{Horowitz:2021dmr}%
  \BibitemOpen
  \bibfield  {author} {\bibinfo {author} {\bibfnamefont {W.}~\bibnamefont
  {Horowitz}}\ and\ \bibinfo {author} {\bibfnamefont {A.}~\bibnamefont
  {Rothkopf}},\ }\href {\doibase 10.21468/SciPostPhysProc.10.025} {\bibfield
  {journal} {\bibinfo  {journal} {SciPost Phys. Proc.}\ }\textbf {\bibinfo
  {volume} {10}},\ \bibinfo {pages} {025} (\bibinfo {year} {2022})},\ \Eprint
  {http://arxiv.org/abs/2109.01422} {arXiv:2109.01422 [hep-ph]} \BibitemShut
  {NoStop}%
\bibitem [{\citenamefont {Umeda}\ \emph {et~al.}(2009)\citenamefont {Umeda},
  \citenamefont {Ejiri}, \citenamefont {Aoki}, \citenamefont {Hatsuda},
  \citenamefont {Kanaya}, \citenamefont {Maezawa}, \citenamefont {Ohno},
  \citenamefont {Collaboration} \emph {et~al.}}]{umeda2009fixed}%
  \BibitemOpen
  \bibfield  {author} {\bibinfo {author} {\bibfnamefont {T.}~\bibnamefont
  {Umeda}}, \bibinfo {author} {\bibfnamefont {S.}~\bibnamefont {Ejiri}},
  \bibinfo {author} {\bibfnamefont {S.}~\bibnamefont {Aoki}}, \bibinfo {author}
  {\bibfnamefont {T.}~\bibnamefont {Hatsuda}}, \bibinfo {author} {\bibfnamefont
  {K.}~\bibnamefont {Kanaya}}, \bibinfo {author} {\bibfnamefont
  {Y.}~\bibnamefont {Maezawa}}, \bibinfo {author} {\bibfnamefont
  {H.}~\bibnamefont {Ohno}}, \bibinfo {author} {\bibfnamefont {W.-Q.}\
  \bibnamefont {Collaboration}},  \emph {et~al.},\ }\href@noop {} {\bibfield
  {journal} {\bibinfo  {journal} {Physical review D}\ }\textbf {\bibinfo
  {volume} {79}},\ \bibinfo {pages} {051501} (\bibinfo {year}
  {2009})}\BibitemShut {NoStop}%
\bibitem [{\citenamefont {Fern{\'a}ndez}\ \emph {et~al.}(2014)\citenamefont
  {Fern{\'a}ndez}, \citenamefont {Hicken},\ and\ \citenamefont
  {Zingg}}]{fernandez2014review}%
  \BibitemOpen
  \bibfield  {author} {\bibinfo {author} {\bibfnamefont {D.~C. D.~R.}\
  \bibnamefont {Fern{\'a}ndez}}, \bibinfo {author} {\bibfnamefont {J.~E.}\
  \bibnamefont {Hicken}}, \ and\ \bibinfo {author} {\bibfnamefont {D.~W.}\
  \bibnamefont {Zingg}},\ }\href@noop {} {\bibfield  {journal} {\bibinfo
  {journal} {Computers \& Fluids}\ }\textbf {\bibinfo {volume} {95}},\ \bibinfo
  {pages} {171} (\bibinfo {year} {2014})}\BibitemShut {NoStop}%
\bibitem [{\citenamefont {Al-Hashimi}\ and\ \citenamefont
  {Wiese}(2021)}]{Al-Hashimi:2021uhu}%
  \BibitemOpen
  \bibfield  {author} {\bibinfo {author} {\bibfnamefont {M.~H.}\ \bibnamefont
  {Al-Hashimi}}\ and\ \bibinfo {author} {\bibfnamefont {U.~J.}\ \bibnamefont
  {Wiese}},\ }\href {\doibase 10.1103/PhysRevResearch.3.L042008} {\bibfield
  {journal} {\bibinfo  {journal} {Phys. Rev. Res.}\ }\textbf {\bibinfo {volume}
  {3}},\ \bibinfo {pages} {L042008} (\bibinfo {year} {2021})}\BibitemShut
  {NoStop}%
\bibitem [{\citenamefont {Albrecht}\ \emph {et~al.}(2022)\citenamefont
  {Albrecht}, \citenamefont {Herrmann}, \citenamefont {Mariani}, \citenamefont
  {Wiese},\ and\ \citenamefont {Wyss}}]{Albrecht:2022sdd}%
  \BibitemOpen
  \bibfield  {author} {\bibinfo {author} {\bibfnamefont {I.}~\bibnamefont
  {Albrecht}}, \bibinfo {author} {\bibfnamefont {J.}~\bibnamefont {Herrmann}},
  \bibinfo {author} {\bibfnamefont {A.}~\bibnamefont {Mariani}}, \bibinfo
  {author} {\bibfnamefont {U.~J.}\ \bibnamefont {Wiese}}, \ and\ \bibinfo
  {author} {\bibfnamefont {V.}~\bibnamefont {Wyss}},\ }\href@noop {} {\
  (\bibinfo {year} {2022})},\ \Eprint {http://arxiv.org/abs/2206.07531}
  {arXiv:2206.07531 [quant-ph]} \BibitemShut {NoStop}%
\bibitem [{\citenamefont {Carpenter}\ \emph {et~al.}(1994)\citenamefont
  {Carpenter}, \citenamefont {Gottlieb},\ and\ \citenamefont
  {Abarbanel}}]{carpenter1994time}%
  \BibitemOpen
  \bibfield  {author} {\bibinfo {author} {\bibfnamefont {M.~H.}\ \bibnamefont
  {Carpenter}}, \bibinfo {author} {\bibfnamefont {D.}~\bibnamefont {Gottlieb}},
  \ and\ \bibinfo {author} {\bibfnamefont {S.}~\bibnamefont {Abarbanel}},\
  }\href@noop {} {\bibfield  {journal} {\bibinfo  {journal} {Journal of
  Computational Physics}\ }\textbf {\bibinfo {volume} {111}},\ \bibinfo {pages}
  {220} (\bibinfo {year} {1994})}\BibitemShut {NoStop}%
\bibitem [{\citenamefont {Lundquist}\ and\ \citenamefont
  {Nordstr{\"o}m}(2014)}]{lundquist2014sbp}%
  \BibitemOpen
  \bibfield  {author} {\bibinfo {author} {\bibfnamefont {T.}~\bibnamefont
  {Lundquist}}\ and\ \bibinfo {author} {\bibfnamefont {J.}~\bibnamefont
  {Nordstr{\"o}m}},\ }\href@noop {} {\bibfield  {journal} {\bibinfo  {journal}
  {Journal of Computational Physics}\ }\textbf {\bibinfo {volume} {270}},\
  \bibinfo {pages} {86} (\bibinfo {year} {2014})}\BibitemShut {NoStop}%
\bibitem [{\citenamefont {Rothkopf}\ and\ \citenamefont
  {Nordstr\"om}(2022)}]{Rothkopf:2022zfb}%
  \BibitemOpen
  \bibfield  {author} {\bibinfo {author} {\bibfnamefont {A.}~\bibnamefont
  {Rothkopf}}\ and\ \bibinfo {author} {\bibfnamefont {J.}~\bibnamefont
  {Nordstr\"om}},\ }\href@noop {} {\  (\bibinfo {year} {2022})},\ \Eprint
  {http://arxiv.org/abs/2205.14028} {arXiv:2205.14028 [math.NA]} \BibitemShut
  {NoStop}%
\bibitem [{\citenamefont {Alvestad}\ \emph {et~al.}(2021)\citenamefont
  {Alvestad}, \citenamefont {Larsen},\ and\ \citenamefont
  {Rothkopf}}]{Alvestad:2021hsi}%
  \BibitemOpen
  \bibfield  {author} {\bibinfo {author} {\bibfnamefont {D.}~\bibnamefont
  {Alvestad}}, \bibinfo {author} {\bibfnamefont {R.}~\bibnamefont {Larsen}}, \
  and\ \bibinfo {author} {\bibfnamefont {A.}~\bibnamefont {Rothkopf}},\ }\href
  {\doibase 10.1007/JHEP08(2021)138} {\bibfield  {journal} {\bibinfo  {journal}
  {JHEP}\ }\textbf {\bibinfo {volume} {08}},\ \bibinfo {pages} {138} (\bibinfo
  {year} {2021})},\ \Eprint {http://arxiv.org/abs/2105.02735} {arXiv:2105.02735
  [hep-lat]} \BibitemShut {NoStop}%
\bibitem [{\citenamefont {Galley}(2013)}]{galley_classical_2013}%
  \BibitemOpen
  \bibfield  {author} {\bibinfo {author} {\bibfnamefont {C.~R.}\ \bibnamefont
  {Galley}},\ }\href {\doibase 10.1103/PhysRevLett.110.174301} {\bibfield
  {journal} {\bibinfo  {journal} {Physical Review Letters}\ }\textbf {\bibinfo
  {volume} {110}},\ \bibinfo {pages} {174301} (\bibinfo {year} {2013})},\
  \bibinfo {note} {publisher: American Physical Society}\BibitemShut {NoStop}%
\bibitem [{\citenamefont {Berges}\ and\ \citenamefont
  {Gasenzer}(2007)}]{berges2007quantum}%
  \BibitemOpen
  \bibfield  {author} {\bibinfo {author} {\bibfnamefont {J.}~\bibnamefont
  {Berges}}\ and\ \bibinfo {author} {\bibfnamefont {T.}~\bibnamefont
  {Gasenzer}},\ }\href@noop {} {\bibfield  {journal} {\bibinfo  {journal}
  {Physical Review A}\ }\textbf {\bibinfo {volume} {76}},\ \bibinfo {pages}
  {033604} (\bibinfo {year} {2007})}\BibitemShut {NoStop}%
\end{thebibliography}%

\end{document}